\documentclass[sigconf]{acmart}

\usepackage{amsmath}     

\usepackage{amssymb}     
\usepackage{amsthm}      
\usepackage{bm}          

\usepackage{enumitem}    
\usepackage{graphicx}    
\usepackage{rotating}    
\usepackage{soul}        

\usepackage{booktabs}    
\usepackage{multirow}    
\usepackage{makecell}    
\usepackage{arydshln}    
\usepackage{tablefootnote} 
\usepackage{diagbox}     
\usepackage{caption}     

\usepackage[linesnumbered,ruled,vlined]{algorithm2e} 

\theoremstyle{plain}

\usepackage{hyperref}

\AtBeginDocument{%
  \providecommand\BibTeX{{%
    Bib\TeX}}}

\copyrightyear{2025}
\acmYear{2025}
\setcopyright{acmlicensed}
\acmConference[KDD '25] {Proceedings of the 31st ACM SIGKDD Conference on Knowledge Discovery and Data Mining V.2}{August 3--7, 2025}{Toronto, ON, Canada.}
\acmBooktitle{Proceedings of the 31st ACM SIGKDD Conference on Knowledge Discovery and Data Mining V.2 (KDD '25), August 3--7, 2025, Toronto, ON, Canada}
\acmISBN{979-8-4007-1454-2/25/08}
\acmDOI{10.1145/3711896.3737266}

\begin{document}

\title{TayFCS: Towards Light Feature Combination Selection for Deep Recommender Systems}

\author{Xianquan Wang}
\affiliation{%
  \institution{University of Science and Technology of China}
  \city{Hefei}
  \country{China}
}
\email{wxqcn@mail.ustc.edu.cn}

\author{Zhaocheng Du}
\authornote{Corresponding author}
\affiliation{%
  \institution{Huawei Noah’s Ark Lab}
  \city{Shenzhen}
  \country{China}
}
\email{zhaochengdu@huawei.com}

\author{Jieming Zhu}
\affiliation{%
  \institution{Huawei Noah’s Ark Lab}
  \city{Shenzhen}
  \country{China}
}
\email{jamie.zhu@huawei.com}

\author{Chuhan Wu}
\affiliation{%
  \institution{Huawei Noah’s Ark Lab}
  \city{Beijing}
  \country{China}
}
\email{wuchuhan1@huawei.com}

\author{Qinglin Jia}
\affiliation{%
  \institution{Huawei Noah’s Ark Lab}
  \city{Beijing}
  \country{China}
}
\email{jiaqinglin2@huawei.com}

\author{Zhenhua Dong}
\affiliation{%
  \institution{Huawei Noah’s Ark Lab}
  \city{Shenzhen}
  \country{China}
}
\email{dongzhenhua@huawei.com}

\renewcommand{\shortauthors}{Xianquan Wang et al.}
\def\BibTeX{{\rm B\kern-.05em{\sc i\kern-.025em b}\kern-.08em
    T\kern-.1667em\lower.7ex\hbox{E}\kern-.125emX}}

\begin{abstract}
Feature interaction modeling is crucial for deep recommendation models. A common and effective approach is to construct explicit feature combinations to enhance model performance. However, in practice, only a small fraction of these combinations are truly informative. Thus it is essential to select useful feature combinations to reduce noise and manage memory consumption. While feature selection methods have been extensively studied, they are typically limited to selecting individual features. Extending these methods for high-order feature combination selection presents a significant challenge due to the exponential growth in time complexity when evaluating feature combinations one by one.
In this paper, we propose \textbf{TayFCS}, a lightweight feature combination selection method that significantly improves model performance. 
Specifically, we propose the Taylor Expansion Scorer (TayScorer) module for field-wise Taylor expansion on the base model. Instead of evaluating all potential feature combinations' importance by repeatedly running experiments with feature adding and removal, this scorer only needs to approximate the importance based on their sub-components' gradients. This can be simply computed with one backward pass based on a trained recommendation model.
To further reduce information redundancy among feature combinations and their sub-components, we introduce Logistic Regression Elimination (LRE), which estimates the corresponding information gain based on the model prediction performance.
Experimental results on three benchmark datasets validate both the effectiveness and efficiency of our approach. 
Furthermore, online A/B test results demonstrate its practical applicability and commercial value. \footnote{For basic concepts of the symbols and terminologies like \textbf{field}, \textbf{feature} and \textbf{combination}, please refer to Appendix~\ref{sec:symbol}.}

\end{abstract}

\ccsdesc[500]{Information systems~Online advertising}

\keywords{Feature Combination; Deep Recommender Systems; Online AD}

\maketitle

\section{Introduction}
\begin{figure}[!t]
    \centering
    \includegraphics[width=1\linewidth]{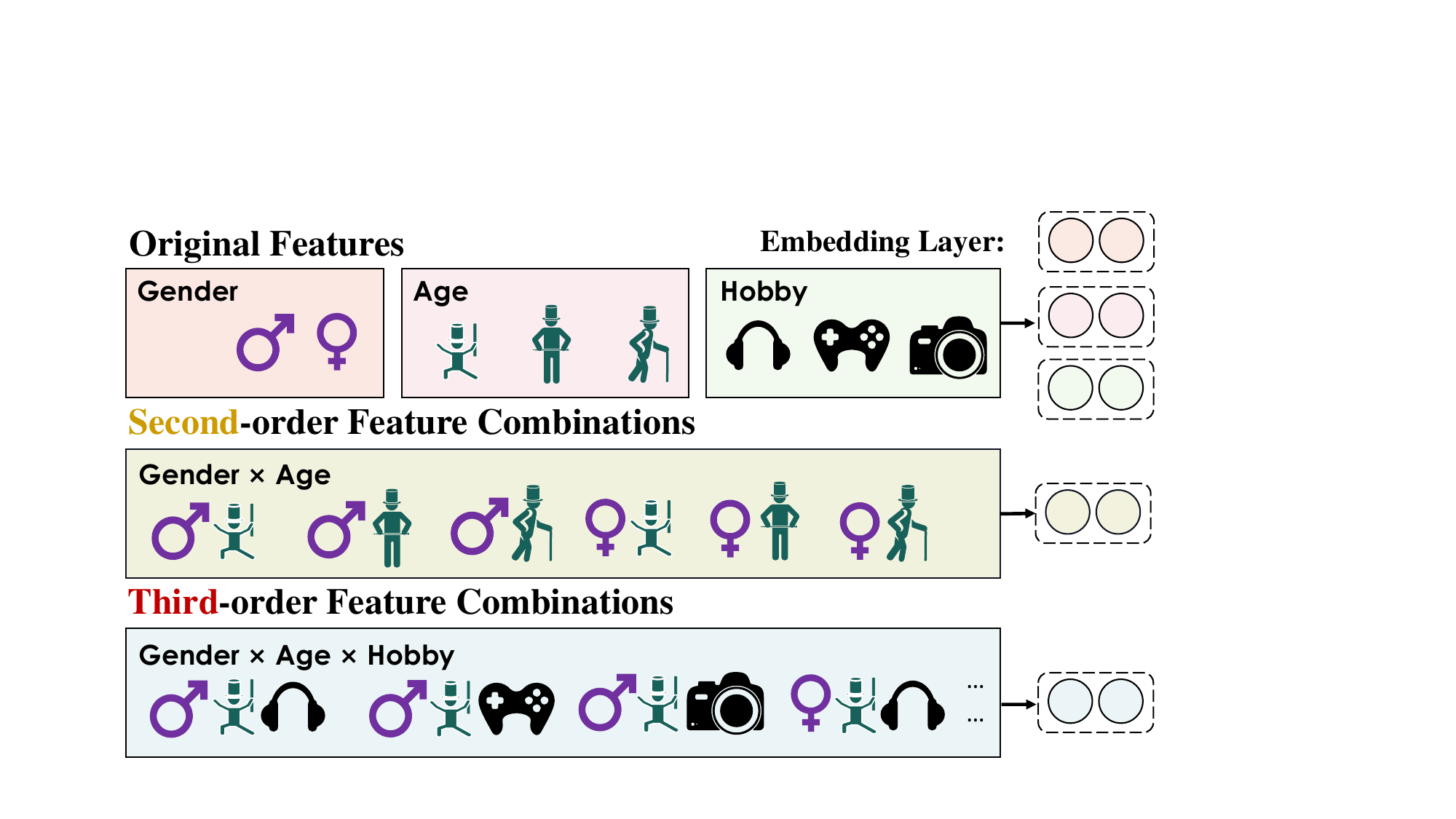}
    \caption{A simple illustration for feature combinations with different orders. The higher-order features are obtained by taking the Cartesian product of the original features.}
    \label{fig:toy}
\end{figure}

Modeling complex user-item interaction behaviors is usually critical in Deep Recommender Systems (DRSs)~\cite{mjyin,da2020recommendation}. 
Due to the powerful pattern recognition abilities of deep neural networks, researchers have developed various methods (\textit{e.g.}, DeepFM~\cite{guo2017deepfm}, DCN~\cite{wang2017deep} and PNN~\cite{qu2016product}) to model the implicit interactions among input features based only on the guidance of supervision signals. 
However, the limitation of model capacity and data amount may hinder these methods from capturing fine-grained and high-order feature interactions based on individual input features~\cite{lyu2022memorize, du2024lightcs}.
Thus, a common industry practice is to explicitly combine different features using Cartesian product~\cite{luo2019autocross} and incorporate them into additional features to reduce the difficulty of modeling feature co-occurrence patterns (see Figure~\ref{fig:toy})~\cite{yao2020efficient,wang2021tedic}.  
However, the exponential growth of feature combinations not only leads to the explosion of computational complexity but also introduces heavy information redundancy to model learning. 
Therefore, selecting a subset of effective feature combinations becomes a critical problem for practical recommender systems. 

Unfortunately, little attention has been paid to exploring feature combination selection in current research. 
The simplest approach is adapting well-researched single-feature selection methods for this task. For example, sensitivity-based methods~\cite{wang2023single} approximate loss change after removing features. Gating-based methods~\cite{wang2022autofield,lin2022adafs} assign trainable gates to each feature and determine feature importance based on the produced gating value. 
However, these methods require pre-constructing all feature combinations and storing them as additional single features, whose memory and time complexity increase exponentially. 
This makes the training and selection process not scalable for practical use.
Other methods have adopted the idea of modeling user-item interaction based on each record by using an interaction module but are still used for selecting single features. For example, 
AdaFS~\cite{lin2022adafs} recognizes the importance of feature interactions, designs controllers to learn the interaction between each feature field, and selects the most important features for each record. \citet{lee2023mvfs} inherit the ideas above by using controllers from multiple views. However, while these methods acknowledge the merit of feature interactions, they still decompose the feature modeling process into individual features, leading to significant storage or computation overhead.

Consequently, current methods poses two challenges. Firstly, the \underline{efficiency bottleneck} of precise feature combination selection methods is introduced by the repeatedly running experiments with feature adding, removal and re-training on a large amount of data.
Secondly, the \underline{feature redundancy issue} is exposed because feature selection methods cannot distinguish information overlap between features and lead to suboptimal feature selection results. This challenge is especially severe between feature combinations and their sub-components because of their dependency relationship. As a result, the industry still lacks an efficient algorithm for selecting important, non-redundant feature combinations that enhance model performance with minimal new feature combinations.

To overcome the challenges mentioned above, in this paper, we propose a novel and efficient \textbf{Tay}lor-based \textbf{F}eature \textbf{C}ombination \textbf{S}election framework. 
Inspired by the fact that the importance signals of feature combinations are hidden in model gradients, enumerating all feature combinations may not be necessary~\cite{schnake2021higher, li2025mergenet}.
We model the relationship between features and the final prediction outcome from the perspective of Taylor expansion, thereby capturing feature importance scores without the need for pre-constructing combined features. Traditional Taylor expansions typically require calculating joint partial derivatives between pairs or triples of feature embeddings~\cite{paudel2022higher}, which leads to high computational complexity~\cite{cooper2020efficient}. To alleviate this, while preserving their rank conditions, we develop an efficient approximation method that rapidly measures combined feature importance~\cite{abbas2021power} with the extension of Information Matrix Equality. Based on the approximated Taylor expansion, this analysis can be applied to any deep recommender system without the need for additional adaptation. 
With the obtained scores, we propose using a simple surrogate model—logistic regression—to eliminate redundant feature combinations. The features are sorted in descending order of importance and added to the surrogate model in batches according to the window size. During validation, features that do not provide positive gains are removed until the added features contribute to information gain.
In summary, our contributions are as follows:
\begin{itemize}[leftmargin=*]
  \item  We propose the TayFCS framework for feature combination selection, which is the first work to analyze feature importance of high-order feature combinations in an effective approximation way.
  \item  We propose a redundancy eliminator that efficiently removes redundant feature combinations using an approximation score from a simple surrogate model.
  \item 
  We conduct extensive experiments on three public benchmarks to show that TayFCS leads to significant performance gains with high efficiency. The comprehensive analysis reveals the necessity 
 and robustness of our framework.
\end{itemize}

Our source code is available at \url{https://github.com/xqwustc/TayFCS}.

\section{Related Work}
\subsection{Deep Recommender Systems (DRSs)}
DRSs present items that best match a user's preferences~\cite{liu2023deep}, and various advanced models have been proposed to better capture user interests and recommend more suitable items. \textbf{DNN}~\cite{covington2016deep} builds interactions with feature embeddings through MLPs. DCN~\cite{wang2017deep} constructs interactions with its cross net. 
\textbf{Wide \& Deep}~\cite{cheng2016wide} uses deep networks and logistic regression for prediction, which enhances both the model's generalization and memorization capabilities.
Moreover, \textbf{DeepFM}~\cite{guo2017deepfm} differs from DCN in that its interaction part is FM, which is used to handle second-order signals.
Some works explored more \textit{physically meaningful} interaction methods. For example, IPNN~\cite{qu2016product} assumes inner product could better model the data, while OPNN uses outer product. 
xDeepFM~\cite{lian2018xdeepfm} assumes its Compositional Feature Interaction module can fully capture feature relations.
MaskNet~\cite{wang2021masknet} provides a new paradigm for feature interaction selection in a serialized and parallelized manner.
However, the flaws are they rely on specific assumptions about \textit{how features interact}~\cite{runlongctr}, which are not universal.

\subsection{Feature Combination and Selection}
There are very few works that directly focus on selecting higher-order combined features.
Most work focuses on single feature selection~\cite{jia2024erase}, and some of them can be extended to high-order features. For example, AutoField~\cite{wang2022autofield} uses a gating mechanism to simulate selection and dropout processes on directed acyclic graphs. 
AdaFS~\cite{lin2022adafs} no longer learns the same feature field importance for all records, but instead learns different importances for each record with a controller. While it considers feature interactions, it cannot explicitly select feature combinations and is still limited to single features.
MvFS~\cite{lee2023mvfs} is an enhanced version of AdaFS. It uses multiple controllers (as different views) to learn for each record and improves the selection logic, which focuses on reweighting features at the record level.
Some of these methods have also been adopted for multi-task settings~\cite{singledu}, but extending them into higher-order settings is still costly and less efficient. TayFCS achieves this from Taylor expansion views, which selects combined features in a practical way with both efficiency and effectiveness.

\section{Preliminaries and Problem Setup}
\subsection{The Basic Process of DRSs}
For DRSs, the data consists of $F$ fields can be represented as ${\mathcal D}=(\boldsymbol{X},\boldsymbol{y}) = ([\boldsymbol{x}_{1},\ldots,\boldsymbol{x}_{i},\ldots,\boldsymbol{x}_{F}],\boldsymbol{y})$,
where $F$ is the number of fields, \( \boldsymbol{x}_{i} \) denotes the one-hot encoding of the feature corresponding to the \( i \)-th field, $\boldsymbol{X}$ consists of each \( \boldsymbol{x}_{i} \) and \( \boldsymbol{y} \in \{1,0\}^{\lvert \mathcal{D} \rvert} \) represents the actual labels for all records.

In most deep recommender systems (DRSs), one-hot features are transformed into low-dimensional embedding vectors through embedding layers~\cite{miao2022het,wang2023towards}. For each input feature \( \boldsymbol{x}_i \), the corresponding embedding is computed as \( \boldsymbol{e}_i = \boldsymbol{V}_i \boldsymbol{x}_i \), where \( \boldsymbol{V}_i \) denotes the embedding table. These embeddings, \( \boldsymbol{e}_1, \boldsymbol{e}_2, \ldots, \boldsymbol{e}_F \), are then concatenated to form the final embedding representation, as \( \boldsymbol{E} = [\boldsymbol{e}_1, \ldots, \boldsymbol{e}_F] \).
For notational convenience in subsequent sections, we reuse $x_i$ to denote the embedding vector $\boldsymbol{e}_i$. Both symbols represent the latent representation of the feature from the $i$-th field.

Then, the embedding representation $\boldsymbol{E}$ will be fed into DRSs~\cite{zhang2021multi,huang2017ad} such as DCN~\cite{wang2017deep} and DeepFM~\cite{guo2017deepfm}. In these models, $\boldsymbol{E}$ performs multi-order interaction processing, followed by linear transformations~\cite{NI2024101439, wang2023cl4ctr} and the activation function~\cite{lin2023map}. Finally, $\hat{\boldsymbol{y}}$ represents the prediction output of the model. For descriptive convenience, the transformation for $\boldsymbol{E}$ of the DRSs is expressed by function $f$, thus we have:
$\hat{\boldsymbol{y}} = f(\boldsymbol{E}; \Theta)$, where $\Theta$ represents parameters of DRSs and is learnable during the training process.

The Binary Cross-Entropy (BCE) loss function~\cite{bishop1995neural} is adopted for the model training, with $\hat{y}_i$ and $y_i$ refer to the click probability and the true label.
The specific form of the loss for DRSs can be expressed as:
$\mathcal{L}(\hat{\boldsymbol{y}},\boldsymbol{y}) = -\frac{1}{|{\mathcal D}|} \sum_{i=1}^{|{\mathcal D}|} \left[ y_i \ln(\hat{y}_i) + (1 - y_i) \ln(1 - \hat{y}_i) \right]$.
\begin{figure*}[!t]
    \centering
    \includegraphics[width=1\linewidth]{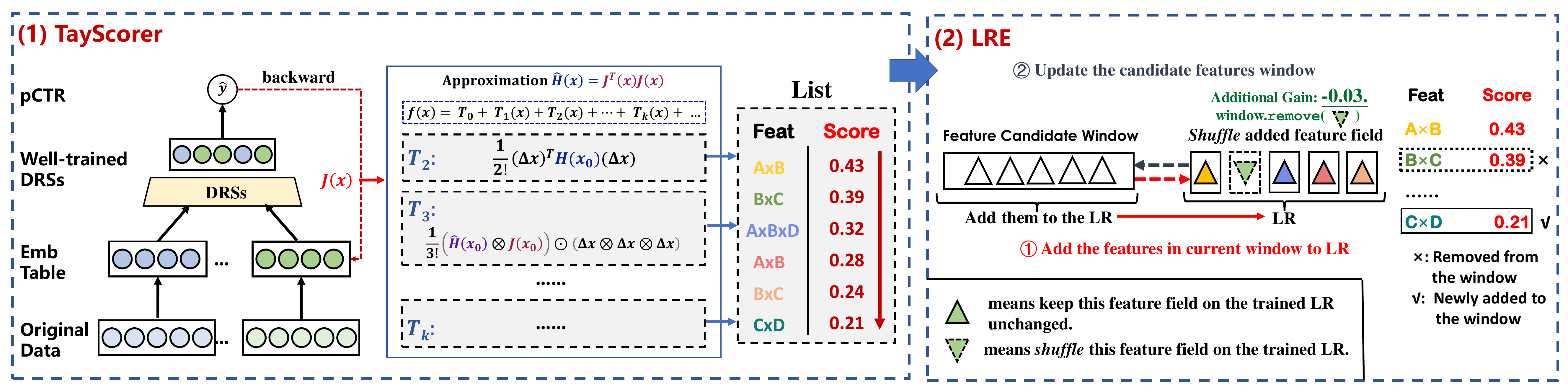}
    \vspace{-2.0em}
    \caption{Overview of the TayFCS framework. First, the TayScorer method in (1) forms a feature list, where features are ranked in descending order based on their importance scores. Then, LRE is used to remove redundant features. The feature colors in (2) correspond to the feature names in (1). This overview shows an example: if some features like \texttt{B×C} are detected by LRE to bring negative gain, they will be removed from the window of candidate features, and the next important features in the feature list are shifted into the feature window.
}
    \label{fig:framework}
    \vspace{-1.0em}
\end{figure*}

\subsection{Feature Combination Selection for DRSs}
Combined Feature Selection involves selecting the most informative subset from candidate feature combinations. In DRSs, these features are treated equally with original features, input directly into the model's input layer, and encoded separately. The primary goal of feature selection is to minimize model loss.
Formally, for a set of feature fields $\mathcal{F}$, we define all its features and combination ones as $\mathcal{P}(\mathcal{F})$, which is also known as the power set (except empty set) of $\mathcal{F}$. Then the candidate feature combinations would be:
\begin{equation}
    \mathcal{P}^*(\mathcal{F})=\mathcal{P}(\mathcal{F})\setminus\ \mathcal{F}.
\end{equation}

We could easily get $|\mathcal{P}^*(\mathcal{F})| = 2^{|\mathcal{F}|} - 1 - |\mathcal{F}|$. 
The ultimate goal is to select a subset of $\mathcal{P}^*(\mathcal{F})$ that enables the model to best learn the feature interaction patterns and achieve optimal prediction performance. During training, we update the model weights $\Theta$ using the cross-entropy loss function as:
\begin{equation}
        \underset{\mathcal{T} \subseteq \mathcal{P}^*(\mathcal{F}), \Theta}{\arg\min} \ \mathcal{L}{(f_{\Theta}(\boldsymbol{X} | (\mathcal{T} \cup \mathcal{F})),\boldsymbol{y})},
\end{equation}
where $\mathcal{T}$ is the expected feature combinations set, $f_{\Theta}$ means the whole DRSs process, and $|$ means using the given features.

\section{Method}
This section presents our feature combination selection framework, which consists of two main parts: efficient analysis of feature importance and elimination of redundant features. The first part is the fast \textbf{Tay}lor Expansion \textbf{Scorer} (TayScorer) module, which analyzes feature importance field-by-field using Taylor expansion. The second part is the \textbf{L}ogistic \textbf{R}egression \textbf{E}limination (LRE) module, which employs a simple logistic regression model to iteratively remove redundant combined features in a sliding feature window, ensuring that the final features are both informative and non-redundant.

\subsection{TayScorer}
\subsubsection{Fast Expansion Method}
We first formalize the relationship between deep models $f$ and features $\boldsymbol{x}$. Here $\boldsymbol{x} = \{x_1, x_2, \dots, x_F\}$ represents the $F$ input features.
Therefore, the $f: \mathbb{R}^F \to \mathbb{R}$ will be a vector function (that means, the input is a vector)\footnote{The output could also be a vector according to the additivity of the total differential. In our task, it is a real number.}.
Based on the above definition, we perform a Taylor expansion on $F$ as:
\begin{equation}
    f(\boldsymbol{x}) = T_0 + T_1(\boldsymbol{x}) + T_2(\boldsymbol{x}) + \dots + T_k(\boldsymbol{x}) + \dots,
    \label{eq:orig}
\end{equation}
where $T_k$ represents the $k$-th order feature combination contribution terms. 
Let 
\begin{equation}
    \boldsymbol{\Delta x} = \boldsymbol{x} - \boldsymbol{x}_0,
\end{equation}
with its components being
\begin{equation}
\Delta{x_i} = x_i - x_{i_0}.    
\end{equation}
With $i = 1,\dots,F$, then we have:
\begin{equation}
    T_k = \frac{1}{k!} \sum_{i_1=1}^F \dots \sum_{i_k=1}^F \frac{\partial^k f(\boldsymbol{x}_0)}{\partial x_{i_1} \dots \partial x_{i_k}} \Delta{x_{i_1}} \dots \Delta{x_{i_k}}.
\end{equation}

Specially, for $T_2$, we have:
\begin{equation}
T_2 = \frac{1}{2} \sum_{i=1}^F \frac{\partial^2 f(\boldsymbol{x}_0)}{\partial x_i^2} \Delta{x_i}^2 + \sum_{i=1}^F \sum_{j=i+1}^F \frac{\partial^2 f(\boldsymbol{x}_0)}{\partial x_i \partial x_j} \Delta{x_i} \Delta{x_j},
\end{equation}
and the higher-order features (like $T_3$) can be found in Appendix~\ref{sec:t3}.
This directly corresponds to the representation with the Hessian matrix, which can be expressed as:
\begin{equation}
T_2 = \frac{1}{2!} (\boldsymbol{\Delta x})^T \mathbf{H}(\boldsymbol{x}_0) (\boldsymbol{\Delta x}),
\label{eq:t2}
\end{equation}
where $\mathbf{H}$ is the Hessian matrix regarding all second-order features. The $\frac{\partial^2f(\boldsymbol{x}_0)}{\partial x_i\partial x_j}\Delta{x_i} \Delta{x_j}$ of $T_2$ is considered as the importance of the given second-order feature combinations.

Intuitively, by performing a Taylor expansion on a well-trained $f$ order by order, \textbf{we can obtain the importance of the corresponding combined features} $(i, j, k, ...)$. However, this approach has an extremely high time complexity, and the exponential growth in the complexity of gradient backpropagation is unacceptable. Therefore, a fast estimation of feature importance is needed.

This estimation of TayFCS is inspired by the perspective of the Information Matrix Equality (IME), a fundamental result in statistical estimation, particularly in maximum likelihood estimation (MLE). IME relates the Hessian (second derivative matrix) of the log-likelihood function to the outer product of its gradient. Specifically, under regularity conditions, it holds that:
\begin{equation}
\mathbb{E} \left[ \frac{\partial^2 \ln f(\boldsymbol{x})}{\partial x_i \partial x_j} \right] = - \mathbb{E} \left[ \frac{\partial \ln f(\boldsymbol{x})}{\partial x_i} \frac{\partial \ln f(\boldsymbol{x})}{\partial x_j} \right],
\label{eq:second}
\end{equation}
which explains the effectiveness of gradient outer product approximations. 
Similarly, the following approximation is employed for the second-order derivatives:
\begin{equation}
    \hat{\mathbf{H}}(\boldsymbol{x}) = \mathbf{J}^T(\boldsymbol{x}) \mathbf{J}(\boldsymbol{x}),
    \label{eq:hes}
\end{equation}
where $\mathbf{J}(\boldsymbol{x}) = \left[ \frac{\partial f(\boldsymbol{x})}{\partial x_1}, \frac{\partial f(\boldsymbol{x})}{\partial x_2}, \frac{\partial f(\boldsymbol{x})}{\partial x_3}, \dots,  \frac{\partial f(\boldsymbol{x})}{\partial x_F}\right]^T$.

Thus, we reduce the gradient backpropagation from $O(n^2)$ to $O(n)$, since multiplying the two gradient vectors $\mathbf{J}(\boldsymbol{x})$ requires only one backpropagation. This is especially reasonable since our loss function is also the BCE with log terms, and we only focus on the \textbf{absolute value} of feature importance as a measure of its contribution.

This allows us to efficiently evaluate the importance of second-order feature combinations. To quickly estimate the importance of higher-order features in a similar way, we heuristically extend this relationship, which enables $O(n)$-time estimation for higher-order feature importance.
Based on this, we could reasonably obtain a third-order joint gradient approximation as:
\begin{equation}
    \hat{\mathbf{H}}(\boldsymbol{x}) \otimes \mathbf{J}(\boldsymbol{x}),
\end{equation}
where $\otimes$ denotes the tensor (outer) product.

For example, the $T_3$ in Equation~\ref{eq:orig} can be expressed as:
\begin{equation}
    T_3 \approx \frac{1}{3!}\left(\hat{\mathbf{H}}(\boldsymbol{x}_0) \otimes \mathbf{J}(\boldsymbol{x}_0)\right) \odot \left(\boldsymbol{\Delta x} \otimes \boldsymbol{\Delta x} \otimes \boldsymbol{\Delta x}\right),
    \label{eq:t3}
\end{equation}
where $\odot$ represents the element-wise production, and each feature combination term in $T_3$ represents its importance.

\subsubsection{The Choice of $\boldsymbol{x}_0$}
In a Taylor expansion, a specific point (also known as the expansion point) must be chosen to approximate the target function. This helps separate the contributions of different feature combinations. In this context, \(\boldsymbol{x}_0\) is the starting point of the Taylor expansion.
 Since we aim to observe the importance of each possible combination feature as accurately as possible through the expansion, this starting point should be as non-informative as possible. Given that each feature's embedding resides in a different feature real number space, we select the mean values of the embeddings for each feature field as \(\boldsymbol{x}_0\) to assess feature contributions more accurately.
This provides different references for the embedding spaces controlled by different feature fields.

Theoretically, we can obtain the importance of features of any order using these estimation methods. Their importance scores represent the contribution to the predicted value of function \( f \). We implemented coarse ranking based on these scores; however, a significant issue is that this approach treats each combination term as independent, which is often not the case in practice. Therefore, we use LRE for further consideration based on the order of importance.

\vspace{-1.0em}
\subsection{LRE}
We hold that adding more features into the network may introduce redundancy. On the one hand, for example, features $\texttt{A}$, $\texttt{B}$, and $\texttt{C}$ might already yield good predictive results when input separately, while their combined one $\texttt{ABC}$ could lead to redundancy. This redundancy can impair the model's prediction accuracy. 
On the other, even some unrelated features may have strong semantic correlations. For example, both $\texttt{postal codes}$ and $\texttt{provinces}$ represent locations. Such informational overlap cannot be directly distinguished by importance scores.
Therefore, we need to estimate whether the newly added feature combinations provide more information gain compared to both the original features and each other.

To address this, we use a greedy algorithm to filter out high-scoring but redundant features. 
\citet{cheng2016wide} pointed out that a linear model has good memorization properties; inspired by this, we adopt Logistic Regression (LR) to assess whether the additional features contribute useful information. As illustrated in Figure~\ref{tb:main}, we utilize a window to consider the additional gain of each candidate feature. The window with size $S_w$ initially contains those features with the highest importance. Each time we add combined features to the logistic regression model, we train the model and then freeze the parameters. As:
\begin{equation} \hat{y} = \sigma \left( \beta_0 + \sum_{i \in (\text{FW}_{S_w} \cup \mathcal{F})} \beta_i x_i \right), \end{equation}
where $\text{FW}$ is the feature window, $S_w$ is the window size, $\beta_i$ is the LR parameters and $\sigma$ is the activation function.

In evaluation, we subsequently shuffle the features of each field $i$ one by one to evaluate the \textbf{additional gain} on model performance. This is based on the validation dataset, as:
\begin{equation}
G(x_i) = \text{Perf}(\text{Feature Window}) - \text{Perf}(\text{Feature Window} \setminus x_i).
\end{equation}

If shuffling some particular fields, on the contrary, improves the model's performance as $G(x_i) \leq 0$ (\textit{i.e.}, these fields brings relatively negative optimization), 
it indicates that some added features are redundant and do not provide any additional benefit. 
In this case, we remove them from the window and continue adding features in descending order, as:
\begin{equation} 
\text{FW}^{(t+1)} = \left( \text{FW}^{(t)} \setminus \{ x_i : G(x_i) \leq 0 \} \right) \cup \{ x_i \in \mathcal{C}_{ls} \},
\label{eq:iter}
\end{equation}
where $\mathcal{C}_{ls}$ is the set of subsequent features added to the feature window in the $(t+1)$-th iteration.
This process will continue until all added features provide additional modeling information ($\forall x_i, G(x_i) > 0$) or the maximum iteration number $T_{\text{iter}}$ (\textit{e.g.}, 1 or 2) is reached. 
In practice, we can use Logloss as an indicator to measure performance changes. If the Logloss value decreases after shuffling a feature, it indicates that the feature is redundant.
Given the rapid training speed of the surrogate model LR, it can effectively eliminates redundancy in practice.

Overall, our algorithm can be represented using the pseudocode in Algorithm~\ref{alg:whole}.

\subsection{Embedding Table Adaption}
Since the added features are combinations of feature fields, which include the product of values, they may contain a large number of values, often requiring substantial resources. To address this, we employ embedding hashing techniques to compress the embedding table. Specifically, for combinations of feature values exceeding a threshold (denoted as $\tau$, set as $5.0e6$), we propose to compress the embedding table to this threshold, and subsequent lookups are performed using the corresponding index obtained via a hash modulo function.

\subsection{Pseudocode of TayFCS}
For clarity and alignment with the equations, in Algorithm~\ref{alg:whole}, we separately present the calculations for second-order (line 3\textasciitilde5) and third-order (line 6\textasciitilde8) feature importance. Higher-order computations can be handled using our proposed approximation method in a similar manner.

\begin{algorithm}
    \caption{The process of TayFCS}
    \label{alg:whole}
    \KwIn{Raw data with full fields $\{ \boldsymbol{x}_i \}_{i=1}^{F}$, the number of added feature combinations $K$, window size $S_w$, maximum feature order $Od_{max}$.}
    \KwOut{$K$ feature combinations in the FW.}
    
    Initialize combined feature importance list: $ls \leftarrow []$. \\
    Train the vanilla model $M$ with original data. \\
    
    \If{$Od_{max} \geq 2$}{
        Calculate all second-order feature importance as Equation~\ref{eq:t2} with approximated Hessian in Equation~\ref{eq:hes}. \\
        Append second-order feature importance to $ls$. \\
    }
    
    \If{$Od_{max} \geq 3$ (Optional)}{
        Calculate all third-order feature importance as Equation~\ref{eq:t3}. \\
        Append third-order feature importance to $ls$. \\
    }
    
    Sort $ls$ in descending order based on the importance score. \\
    $iter \leftarrow 0$ \\
    $\text{Feature Window} \leftarrow ls[:S_w]$. \\

     \While{$iter < T_\text{iter}$ \textbf{and} \textbf{there exists a feature that brings no gain}}{
        $iter \gets iter + 1$ \;
        LR.\texttt{use}(\text{Feature Window}) and train it. \\
        Shuffle each feature field and re-evaluate the gain of each field. \\
        Feature Window.\texttt{remove}(\text{features with no gain}). \\
        Feature Window.\texttt{add}(subsequent features from $ls$). \\
    }
    \Return{$ls[:K]$}
\end{algorithm}

\section{Experiment}
In this section, we explore how the TayFCS framework efficiently and effectively select feature combinations.
Our study focuses on and emphasizes the following questions:
\begin{itemize}[leftmargin=*]
\item \textbf{RQ1:} How does TayFCS perform compared to other recent feature engineering methods for deep recommenders?
\item \textbf{RQ2:} How about the TayScorer results?
\item \textbf{RQ3:} Can the selected combinations of TayFCS be transferred to feature crossing or feature selection models (\textit{e.g.,} \textbf{DCN} and \textbf{MaskNet})?
\item \textbf{RQ4:} How does hyperparameter affect the prediction results?
\item \textbf{RQ5:} How is the efficiency of TayFCS?
\item \textbf{RQ6:} What is the influence of TayFCS core components?
\item \textbf{RQ7:} How about the inference efficiency with additional feature combinations?
\item \textbf{RQ8:} Is the performance of TayFCS consistent between online and offline settings?
\end{itemize}

\begin{table}[!t]\footnotesize
\centering
\renewcommand{\arraystretch}{1.3}
\caption{Statistics of three datasets used for evaluation.}
\begin{tabular}{llllll}
\toprule
\textbf{Dataset} & \textbf{\#Fields} & \textbf{\#Training} & \textbf{\#Validation} & \textbf{\#Test} & \textbf{Positive\%}\\ \hline 
Frappe & 10 &202,027 & 57,722&28,860&33.46\%\\ \hline 
iPinYou & 16 & 13,195,935 & 2,199,323 & 4,100,716 & 0.08\%\\ \hline 
Avazu & 24 & 32,343,172 & 4,042,897 & 4,042,898 & 16.98\%\\ 
\bottomrule
\end{tabular}
\label{tb:dataset}
\end{table}

\subsection{Experiment Settings}
\subsubsection{Dataset}
\label{sec:dataset}
We assess the performance of TayFCS on three public datasets: \textbf{Frappe}, \textbf{iPinYou} and \textbf{Avazu}.
These datasets include large amounts of interaction data collected from real-world scenarios.
Researchers have also claimed that experiments conducted on these datasets are more meaningful to industrial practitioners~\cite{DBLP:conf/cikm/ZhuLYZH21}.

Table~\ref{tb:dataset} shows the number of records in each dataset following the partition ratio. 
Higher-order features require more resources, while second- and third-order features already provide sufficient modeling information. 
For all possible feature combinations, due to their exponential possibilities, we only consider second- and third-order features. In the process of using AutoField (\textit{a.k.a.} AutoField+) to measure feature importance, the batch size of candidate features is set to 20, and the 20 features are pre-constructed and fed into the input layer at a time. Their importance is then learned/measured, and all candidate features are ranked based on their importance.

Following previous research, we divide the Frappe dataset in a \textbf{7:2:1} ratio. For iPinYou, we additionally allocate 1/7 of the training data as a validation set. For Avazu, we use an 8:1:1 split ratio.
\begin{itemize}[leftmargin=*]
\item \textbf{Frappe\footnote{\url{https://www.baltrunas.info/context-aware}}:} This dataset is a collection of context-aware app usage logs, consisting of 96,203 entries from 957 users across 4,082 apps, reflecting diverse usage contexts with 10 feature fields. 
\item \textbf{iPinYou\footnote{\url{https://contest.ipinyou.com/}}:} This dataset is from the iPinYou Global RTB Bidding Algorithm Competition, held in 2013 across three seasons. It includes all training datasets and leaderboard testing datasets for each season, featuring DSP bidding, impression, click, and conversion logs. The reserved testing datasets for final evaluation are withheld by iPinYou.
\item \textbf{Avazu\footnote{\url{https://www.kaggle.com/c/avazu-ctr-prediction/}}:} This dataset contains nearly 40 million interaction records across 22 fields. It has been updated to include 24 fields by splitting the original \textit{hour} field into \textit{weekday}, \textit{weekend}, and a new \textit{hour} field.
\end{itemize}

\subsubsection{Models and Baseline Methods}
\label{sec:baseline}
The backbone models we use are: (i) \textbf{DNN}~\cite{covington2016deep}; (ii) \textbf{DeepFM}~\cite{guo2017deepfm} ; (iii) \textbf{Wide \& Deep}~\cite{cheng2016wide}. Moreover, we use these models for transfer learning on the optimization results: (iv) \textbf{DCN}~\cite{wang2017deep}; (v) \textbf{MaskNet}~\cite{wang2021masknet}. All of them are widely applied or recently proposed models. 

We evaluate the TayFCS with these methods: (i) \textbf{Randomly Select}, which selects the feature combinations from all possible feature candidates; (ii) \textbf{AutoField} \cite{wang2022autofield}, which learns single value importance to perform feature selection at the field level. For comparison, we \textit{extend} \textbf{AutoField} to high-order feature combination candidates, as \textbf{AutoField+}. It also adopts the hash table adopted as TayFCS does; (iii) \textbf{AdaFS}~\cite{lin2022adafs}, which adopts controllers to learn the feature importance at sample-level. This makes
the training and selection process on each interaction; (iv) \textbf{MvFS}~\cite{lee2023mvfs}, which uses controllers from different views to get more stable importance scores.
Note that for a quick comparison, we obtain AutoField+ and TayFCS results \textbf{on DNN}, and use them for all base models.

\subsubsection{Evaluation Metrics}
\setcounter{footnote}{0}
\renewcommand{\thefootnote}{\fnsymbol{footnote}}
Following the previous works~\cite{wang2022autofield,lin2022adafs,lee2023mvfs}, we adopt AUC and Logloss as performance evaluation metrics. 
Moreover, we use the $\Delta$RelImp\footnote{$\Delta$RelImp = $(\frac{\text{AUC(Method)}-0.5}{\text{AUC(Original Model)}-0.5}-1)\times100\%$  \vspace{-1.0em}} to measure the relative importance of each method compared to the original methods.
While AUC and $\Delta$RelImp demonstrate superiority at higher values, and the Logloss exhibits superiority at lower levels. Note that a rise in AUC or a reduction in Logloss of more than \ul{0.001 is regarded as a noteworthy enhancement} in performance~\cite{guo2017deepfm,qu2016product}, and we achieved this only by modifying the input features \textbf{without any modification} to the model \textbf{structure} or its \textbf{hyperparameters}.

\setcounter{footnote}{0}
\renewcommand{\thefootnote}{\arabic{footnote}}

\begin{table*}[t!]
\small
    \renewcommand{\arraystretch}{1.1}
    \centering 
    \caption{Performance comparison of different feature (combination) selection methods. }
    \resizebox{0.87\linewidth}{!}{
\begin{tabular}{cccc ccc ccc  ccc}\toprule
\multirow{2}{*}{Model}&\multirow{2}{*}{Method}&\multicolumn{3}{c}{Frappe}& \multicolumn{3}{c}{iPinYou} &\multicolumn{3}{c}{Avazu}\\ \cline{3-13}
 && AUC$\uparrow$ & Logloss$\downarrow$& $\Delta$RelImp$\uparrow$ & AUC$\uparrow$ & Logloss\%$\downarrow$ & $\Delta$RelImp$\uparrow$  &AUC$\uparrow$& Logloss$\downarrow$ & $\Delta$RelImp$\uparrow$ &\\\hline
 \multirow{6}{*}{DNN}& Original & 0.9735 & 0.1825 & - & 0.7795 & 0.5554 & - & 0.7937 & 0.3726 & -\\
 & Random & 0.9718 & 0.1947 & -0.36\% & 0.7800 & \underline{0.5550} & 0.18\% & 0.7936 & 0.3718 & -0.03\%\\
 & AutoField+ & \underline{0.9865} & \underline{0.1292} & \underline{2.74\%} & 0.7793 & 0.5557 & -0.08\% & \underline{0.8013} & \textbf{0.3671} & \underline{2.59\%}\\
 & AdaFS & 0.9799 & 0.1664 & 1.34\% & 0.7799 & 0.5556 & 0.14\% & 0.7965 & 0.3701 & 0.95\%\\
 & MvFS & 0.9804 & 0.1582 & 1.47\% & \underline{0.7803} & 0.5554 & \underline{0.27\%} & 0.7971 & 0.3696 & 1.18\%\\
 & TayFCS & \textbf{0.9868} & \textbf{0.1262*} & \textbf{2.81\%} & \textbf{0.7807} & \textbf{0.5547} & \textbf{0.43\%} & \textbf{0.8021*} & \underline{0.3672} & \textbf{2.88\%}\\\hline

 \multirow{6}{*}{DeepFM}& Original & 0.9726 & 0.1886 & - & 0.7784 & \underline{0.5551} & -  & 0.7942 & 0.3713 & -\\
 & Random & 0.9711 & 0.1923 & -0.33\% & \underline{0.7811} & 0.5552 & \underline{0.98\%} & 0.7944 & 0.3711 & 0.09\%\\
 & AutoField+ & \underline{0.9848} & \underline{0.1293} & \underline{2.57\%} & 0.7800 & 0.5556 & 0.60\% & \underline{0.7991} & \underline{0.3687} & \underline{1.66\%}\\
 & AdaFS & 0.9724 & 0.2025 & -0.05\% & 0.7795 & 0.5555 & 0.40\% & 0.7962 & 0.3704 & 0.67\%\\
 & MvFS & 0.9751 & 0.2084 & 0.53\% & \textbf{0.7812} & \textbf{0.5549} &\textbf{ 1.01\%} & 0.7966 & 0.3701 & 0.82\%\\
 & TayFCS & \textbf{0.9871*} & \textbf{0.1250*} & \textbf{3.06\%} & 0.7784 & 0.5563 & -0.00\% & \textbf{0.8021*} & \textbf{0.3666*} & \textbf{2.70\%}\\\hline

 \multirow{6}{*}{Wide \& Deep}& Original & 0.9770 & 0.1903 & - & 0.7765 & \textbf{0.5556} & - & 0.7941 & 0.3714 & -\\
 & Random & 0.9712 & 0.1903 & -1.21\% & \textbf{0.7800} & 0.5585 & \textbf{1.24\%} & 0.7947 & 0.3710 & 0.20\%\\
& AutoField+& \underline{0.9857} & \underline{0.1281} & \underline{1.84\%}& 0.7735 & 0.5592 & -1.10\% & \underline{0.8009} & \underline{0.3679} & \underline{2.29\%}\\
 & AdaFS &0.9728 & 0.1911 & -0.87\% &  \underline{0.7793} & 0.5573 & \underline{0.99\%} & 0.7967 & 0.3699 & 0.88\% \\
 & MvFS & 0.9796 & 0.1746 & 0.56\% & 0.7771 & 0.5585 & 0.21\% & 0.7970 & 0.3699 & 0.97\%\\
 & TayFCS & \textbf{0.9866*} & \textbf{0.1279} & \textbf{2.02\%} & 0.7762 & \underline{0.5570} & -0.10\% & \textbf{0.8027*} & \textbf{0.3663*} & \textbf{2.92\%}\\
\bottomrule
\end{tabular}
}
\caption*{\small \normalfont The best result is in \textbf{bold}, and the second best result is \ul{underlined}. The symbols * denotes the t-test significance levels with $p \leq 0.05$ of TayFCS \textbf{compared to the best performing baseline}. 
\textbf{AutoField+} is the higher-order feature version of AutoField. Both \textbf{AutoField+} and Random use the hash embedding table, like TayFCS, for a fair comparison. For all base models, we add 5, 5, and 10 feature combinations for Frappe, iPinYou, and Avazu, respectively.
Since the loss on the iPinYou dataset is small, we use Logloss\% instead of Logloss here.}

\label{tb:main}
\end{table*}

\subsubsection{Hyperparameters}
\label{sec:imple}
This section offers a detailed description of how our experiments were carried out.
We use FuxiCTR repository to conduct our experiments. About our TayFCS framework:

(i) \textbf{General hyperparameters:} 
For all datasets, the batch size is 10,000, and the maximum number of training epochs is set to 100, ensuring that all models are trained satisfactorily.
The early stopping strategy is implemented, wherein premature termination occurs if there is no increase in AUC-Logloss on the validation set for two consecutive epochs.
Standard techniques like the Adam optimizer~\cite{DBLP:journals/corr/KingmaB14} and Xavier initialization~\cite{glorot2010understanding} are also adopted.

(ii) \textbf{TayFCS hyperparameters:} For TayFCS, $T_{\text{iter}}$ is to obtain the importance score, which refers to the maximum iterations of LRE. In our experiments, \textbf{setting it to 1} already brings additional improvement. Moreover, the number of feature combinations added to the model (\textit{a.k.a.} $K$) is a hyperparameter shared by all selection methods, and we seach it within [5, 10, 15, 20]. The maximum order of features  $Od_{max}$ for all selection methods is set to 3, and the window size $S_w$ is searched within [10, 20].

\subsection{Main Performance (RQ1)}
Table~\ref{tb:main} presents the main experimental results, which are conducted on three datasets with three models. From the analysis of the table, we can draw the following conclusions:

First, the impact of feature combinations varies across datasets, which means the performance improvement from adding feature combinations differs significantly. For example, on the Frappe dataset, combined features yield an improvement of 2\textasciitilde3\%, similar to Avazu.
According to the benchmark provided by BARS~\cite{zhu2021open}, TayFCS, based on three DNN-based base models, has achieved a new SOTA on Frappe\footnote{\url{https://github.com/reczoo/BARS/blob/main/docs/CTR/leaderboard/frappe_x1.csv}} and Avazu\footnote{\url{https://github.com/reczoo/BARS/blob/main/docs/CTR/leaderboard/avazu_x4_002.csv}} at the leaderboard by selecting feature combinations.
But the improvement on iPinYou is smaller. The reason for this phenomenon may be that the positive samples in the iPinYou dataset are sparse, and useful information in the form of gradients is relatively limited in the validation set. As a result, the trained model may struggle to distinguish useful combinations. Moreover, due to this distribution characteristic, the model cannot effectively learn combined features during re-training, leading to a relatively small improvement.

Second, even within the same dataset, different models show varying improvements, mainly due to differences in modeling capacity. For example, Frappe performs best on DeepFM, which may be due to its smaller number of fields, allowing FM to capture higher-order interaction information through cross-product on selected feature combinations. On iPinYou, DNN achieves the best performance, likely because other interaction modules struggle to learn due to the sparsity of positive samples. On Avazu, the LR component in Wide \& Deep may provide strong memory capability, which clearly aids the model in learning user patterns.

Third, there are significant differences between selection methods. Randomly selecting feature combinations (Random) from second- and third-order features shows decent results in a few cases, but in most instances, it provides the lowest improvement, and sometimes even causes a significant performance drop (\textit{e.g.}, Frappe on DNN). This indicates that merely adding combined features to the network without considering their informative value does not help the model learn interaction information better. 
AdaFS and MvFS controllers achieve relatively good results on the iPinYou dataset, but their performance is less impressive on Frappe and Avazu. In most cases, AutoField+ and TayFCS achieve top-2 performance, indicating that feature combinations constructed at the input layer help the model learn the complex interaction relationships between users and items, which cannot be achieved solely through changes in the model architecture.

These conclusions clearly demonstrate the strong effectiveness of TayFCS in feature combination selection.

\subsection{Visual Importance Results (RQ2)}
\begin{figure}
    \centering
    \includegraphics[width=1.0\linewidth]{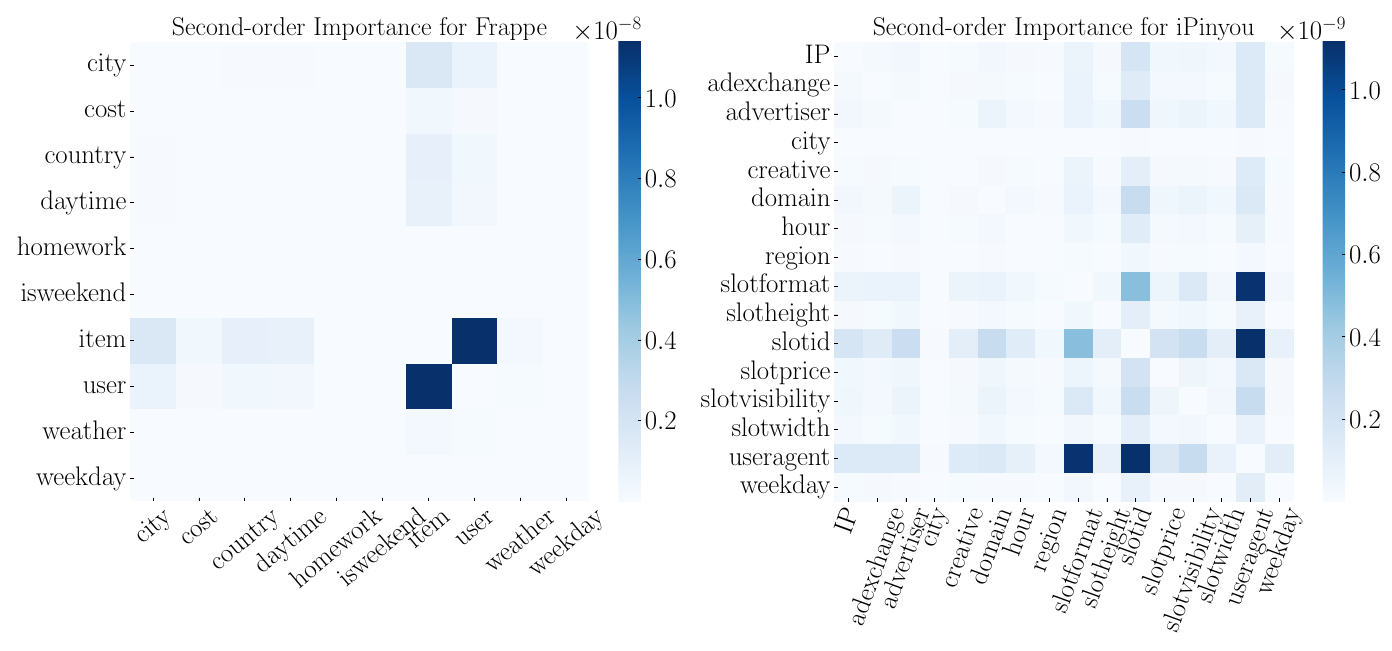}
    \caption{Visual importance results for Frappe and iPinYou dataset.}
    \label{fig:visual}
    \vspace{-0.4cm}
\end{figure}

To intuitively present the combined feature scores we analyzed, we visualize the importance scores using a heatmap. For simplicity, we only show the importance scores for second-order features. Since the real names of feature fields can help us better understand interactions between fields, we display results from the Frappe and iPinYou datasets (on DNN).

As shown in Figure~\ref{fig:visual}, it can be observed that important second-order combined features are sparse, which aligns with the motivation mentioned in the introduction. 
For the Frappe dataset, it is clear that combined features composed of user and item are the most important, which is intuitive. 
Moreover, \texttt{country} and \texttt{daytime} also play an important role, as they help the model capture variations related to the environment and time of the advertisement display.

For the iPinYou dataset, the important features mainly involve features combinations related to \texttt{slotid} and \texttt{useragent}. 
Next are some features related to \texttt{slotvisibility} or \texttt{slotformat}, which indicates that the combination of whether an advertisement slot is visible, its format, and other features generates stronger discriminative signals.

The results show that our method effectively highlights important features compared to less important ones.

\subsection{Transfer Learning (RQ3)} \label{sec:transfer}
\begin{figure}
    \centering
    \includegraphics[width=0.9\linewidth]{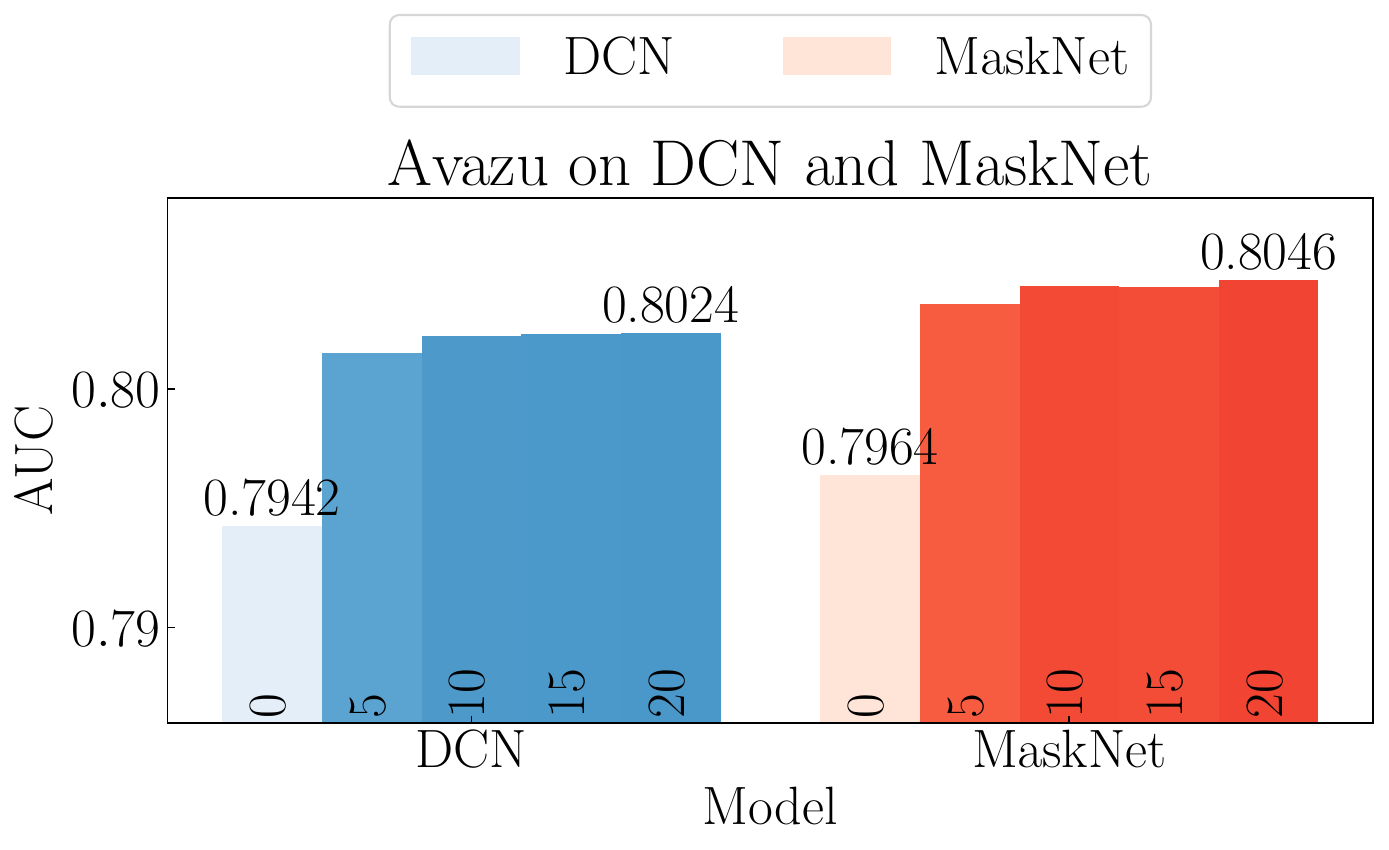}
    \caption{Transfer learning for Avazu dataset. The numbers on the bars in the figure represent the number of added feature combinations, with 0 indicating the original model.}
    \label{fig:transfer}
\end{figure}
Exploring whether the combination selection results can generalize to other models is an important way to assess its robustness. To this end, we use \textbf{DCN and MaskNet as the target model} and \textbf{DNN as the source model}. DCN adopts the well-known crossnet to implicitly model the feature interactions. MaskNet employs an instance-guided mask, performing element-wise multiplication on feature embeddings and feedforward layers. Moreover, MaskNet transforms the feedforward layers in the DNN into a hybrid form that combines additive and multiplicative feature interactions, which aims to fully explore interaction patterns. 
We add various feature combinations to both models to examine \textbf{whether} these models have already sufficiently \textbf{learned the interaction signals}.

As shown in Figure~\ref{fig:transfer}, the feature combinations in DCN and MaskNet further improve performance, an improvement that cannot be achieved by model adjustments alone. 
For DCN, the AUC improved from 0.7942 to 0.8024. On the well-designed MaskNet model (initially 0.7964), the AUC reached 0.8046 after incorporating feature combinations on Avazu, setting a new SOTA for this dataset.
This not only demonstrates the effective transferability of the combination results produced by our method, but also indicates that relying solely on model optimization (such as MaskNet) is insufficient for adequately modeling user-item relationships.
The use of TayFCS features has indeed led to great improvements over the base models.

\subsection{Hyperparameter Analysis (RQ4)}
\begin{figure}
    \centering
    \includegraphics[width=1.0\linewidth]{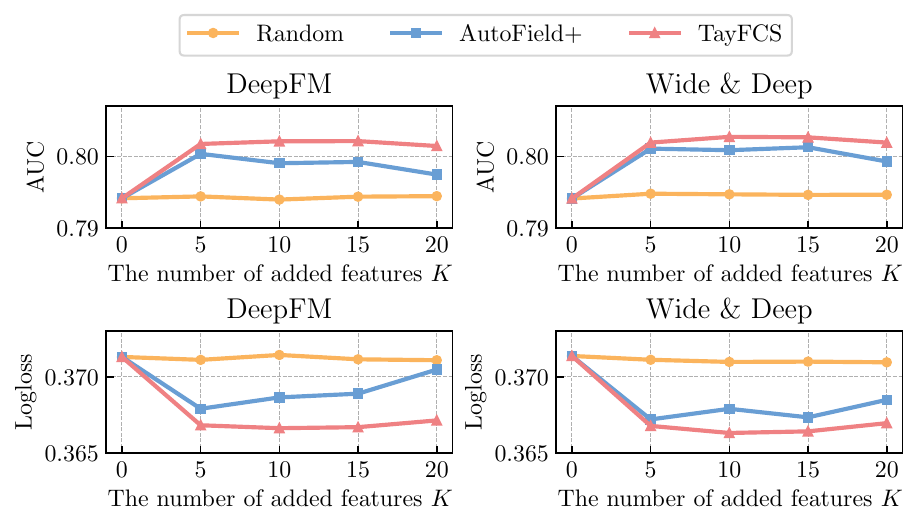}
    \caption{The performance with different numbers of feature combinations $K$ with Avazu dataset on various base models. The upper section shows AUC, while the lower section presents the corresponding Logloss.}
    \label{fig:hyper}
    \vspace{-0.5cm}
\end{figure}
For all explicit selection methods, we examine the relationship between the number of added feature combinations and model performance. We analyze it for Avazu on DeepFM and Wide \& Deep.
As shown in Figure~\ref{fig:hyper}, overall, the TayFCS we proposed (\textcolor{red}{red line}) clearly outperforms other methods. 

TayFCS achieved a significant improvement over the basemodel when $K=5$, and reached the best performance at $K=15$. However, at $K=20$, the model performance slightly declined. The same trend can be observed on Wide \& Deep. This indicates that the informative combinations subset is indeed a small fraction of the features. As more features are added, excessive information may make it difficult for the model to distinguish which features are more important for measuring user behavior, leading to some performance degradation. The performance at $K=10$ and $K=15$ is very similar, and considering the balance between required resources and performance, $K=10$ is generally preferred. Moreover, TayFCS consistently outperforms AutoField+ across different $K$ values, while randomly selecting the combination subset fails to provide significant performance improvement, which further highlights the effectiveness of TayFCS.

Overall, it is vital to select an appropriate number of feature combinations considering resources and prediction performance.

\subsection{Analysis Efficiency (RQ5)}
\begin{figure}
    \centering
    \includegraphics[width=1.0\linewidth]{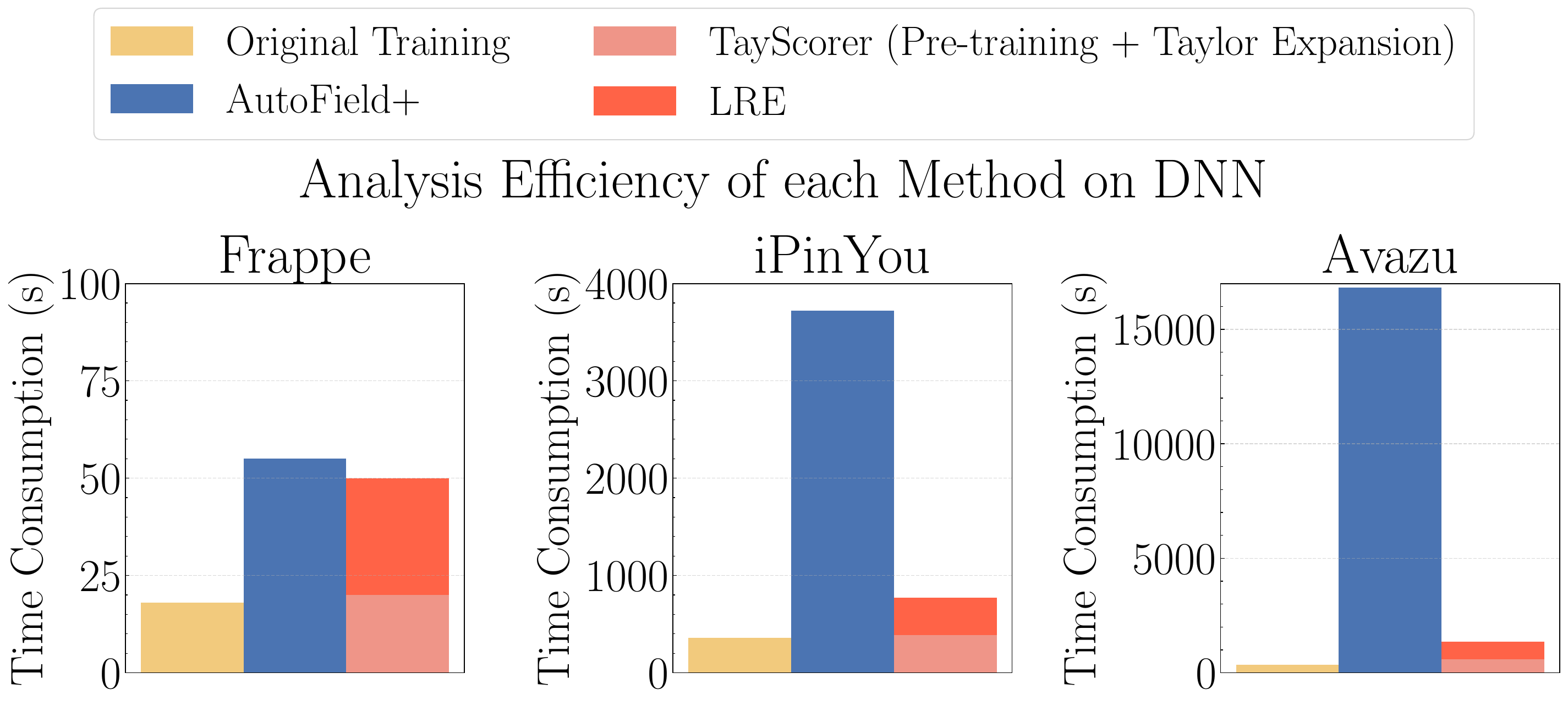}
    \caption{Time consumption for analysis on three datasets. For TayFCS, the time cost is devided into 2 parts: on TayScorer \& LRE.}
    \label{fig:time}
    \vspace{-0.6cm}
\end{figure}
The efficiency of analyzing each feature combination is critical, as excessive time spent on exploring each feature can be resource-intensive and costly. 
In this subsection, we present the \textbf{total time} required to analyze feature combinations, comparing our work with previous feature engineering works.

According to the Experiment Settings, for AutoField+, we group candidate feature combinations into sets of 20 to compute their importance. 
As shown in Figure~\ref{fig:time}, on Frappe, due to its relatively small size, the time consumption of TayFCS is comparable to that of AutoField+. However, on the iPinYou and Avazu datasets, which have more feature fields, TayFCS significantly outperforms AutoField+ in terms of time efficiency. This is because TayFCS employs TayScorer, which approximates higher-order Taylor expansions efficiently. TayScorer estimates higher-order importance scores with a single gradient pass, eliminating the need for multiple training or backpropagation steps. The LRE process only requires training a lightweight logistic regression model once. As a result, its overall time complexity remains similar to that of original training.

\subsection{Ablation Studies (RQ6)}
\begin{table}[t!]\footnotesize
    \centering
    \caption{Ablation study for TayFCS.}
    \renewcommand{\arraystretch}{1.3}
    \resizebox{\linewidth}{!}{
\begin{tabular}{cllllll}\toprule
& \multirow{2}{*}{ Model } & \multirow{2}{*}{ Metric } & \multicolumn{4}{c}{ Method } \\
\cline{4-7} & & & \textbf{\textit{r.s.}} & \textbf{\textit{w.o.h.}} & \textbf{\textit{w.l.r.e.}} & \textbf{TayFCS}\\
\hline
\multirow{4}{*}{\rotatebox{90}{Frappe}} & \multirow{2}{*}{DNN} & AUC $\uparrow$  & 0.9718  & \textbf{0.9868} & 0.9861 & \textbf{0.9868} \\
 & & Logloss $\downarrow$ & 0.1947 & \textbf{0.1262} & 0.1269 & \textbf{0.1262} \\
\cline{2-7}
 & \multirow{2}{*}{DeepFM} & AUC $\uparrow$ & 0.9711  & \textbf{0.9871}  & 0.9848  &  \textbf{0.9871}    \\
 & & Logloss $\downarrow$ & 0.1923  & \textbf{0.1250}  & 0.1293  &  \textbf{0.1250}    \\
\cline{2-7}
 \hline \multirow{4}{*}{\rotatebox{90}{Avazu}} 
 & \multirow{2}{*}{ DNN } & AUC $\uparrow$ & 0.7936  & 0.7937  & 0.8013 & \textbf{0.8021}  \\
 & & Logloss $\downarrow$ & 0.3718 & 0.3715  & 0.3670 &  \textbf{0.3672} \\
\cline{2-7} & \multirow{2}{*}{ DeepFM } & AUC $\uparrow$ & 0.7944  & 0.7946  & 0.8006  & \textbf{0.8021}  \\
  & & Logloss $\downarrow$ & 0.3711  & 0.3712  & 0.3679  &  \textbf{0.3666} \\
  \bottomrule
\end{tabular}}
 {\caption*{\footnotesize \normalfont \textbf{Bold} font indicates the best-performing method result. The experiments of the same group use \textbf{the same number} of featuer combinations.}}
    \label{tb:aba}
    \vspace{-0.6cm}
\end{table}
In the ablation study, we explore whether arbitrarily adding feature combinations can yield good results, whether the hash table plays a role, and how the model behaves when the LRE component is removed. 
They correspond to three ablations: (a) randomly selection (\textbf{\textit{r.s.}}): combined features are selected randomly; (b) without hash table (\textbf{\textit{w.o.h.}}): for combination features, we no longer use a hash table. Instead, combinations exceeding the threshold ($\tau = 5.0e6$) are skipped, and the combinations to be added are determined by feature importance; (c) without LRE (\textbf{\textit{w.l.r.e.}}): removing the LRE component and the combination results obtained using TayFCS are directly fed into the model.
Same as the main experiment, for Frappe, the number of added features is 5, and for Avazu, the number is 10.

From Table~\ref{tb:aba}, we can conclude that each component of TayFCS is crucial. Random feature selection performs the worst, as it is blind and lacks depth. The hash table is essential for models on the Avazu dataset when combinations involve many features.
Without it, performance drops sharply, nearly matching that of random combinations.
However, for Frappe, it is less important because the most important combinations' features do not exceed the threshold $\tau$. Moreover, removing LRE results in a performance drop on both Frappe and Avazu, indicating that LRE effectively removes redundant features. This highlights the effectiveness and adaptability of our proposed TayFCS across different datasets.

\subsection{Inference Time (RQ7)}
\label{sec:inf}
\begin{figure}[t]
    \centering
    \includegraphics[width=1.0\linewidth]{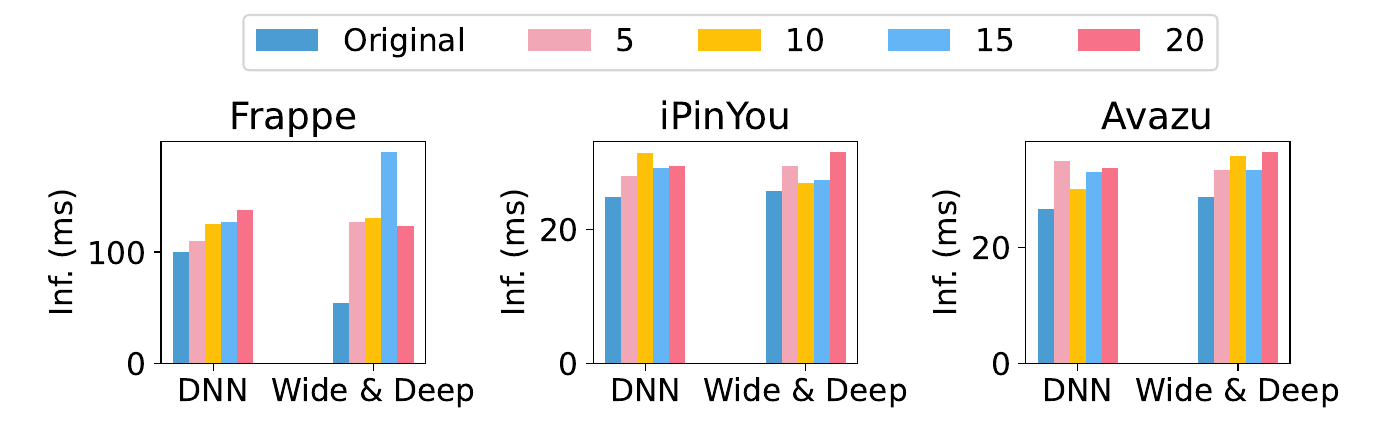}
    \caption{Inference time with three dataset on DNN and Wide \& Deep. The time cost is measured in milliseconds.}
    \label{fig:inf_time}
    \vspace{-0.5cm}
\end{figure}

Inference time is equally important for online services. We evaluate it on the test dataset and report the time in milliseconds \textbf{per batch}.
As shown in Figure~\ref{fig:inf_time}, it is observed that adding feature combinations based on the hash embedding table in TayFCS does not significantly increase inference overhead in most cases. For example, in Frappe on DNN, inference time increased by 10\%, and on iPinYou and Avazu, the increases were 12.5\% and 13.7\%, respectively. A similar trend is observed on Wide \& Deep. 
The increase in time cost mainly comes from two factors: first, the feature combination embedding takes more time for table lookup; second, more combinations require wider hidden states in the model.

Though the added feature combinations account for 50\%, 31.2\%, and 41.7\% of the original feature fields, the increase in inference time is small. This suggests that inference latency does not grow linearly with the number of added features, since the model's computation (excluding embeddings) also takes a large portion of the time. It is also possible that the hash operations used in the embedding table make the lookup overhead relatively low.
Despite the slight increase in time, the improvement in inference performance justifies this additional overhead, as there is \textbf{no free lunch}.

\subsection{Online Test (RQ8)}
We deployed TayFCS in the CVR (Coversion Rate) prediction task on our ad platform. The CVR model is updated daily with billions of samples and hundreds of finely crafted features. To prevent training dataset overfitting, we use 10\% of the validation set (about 50 thousand samples) for TayFCS for feature combination selection. TayFCS takes approximately 1 hour to screen all possible sparse feature combinations and the top 12 features are selected. In our A/B tests, the control group used the TAML~\cite{liu2023task} model with the original features, while the experimental group added 12 combination features from TayFCS (8 2\textsuperscript{nd}-order and 4 3\textsuperscript{rd}-order combinations). Both groups had 5\% traffic from randomly divided users. After two weeks’ observation, the TayFCS group saw a 13.9\% CVR improvement and 0.73\% revenue growth, now serving the whole traffic. TayFCS has become a foundational tool in our platform for feature combination selection, which has resulted in improved user engagement.

\section{Conclusion}
To address the inefficiency and redundancy in current combined feature selection, we propose the TayFCS framework based on Taylor expansion. To speed up the expansion process without sacrificing accuracy, we integrate the IME theory to derive an efficient approximation of Taylor expansion, and then introduce the LRE to reduce the feature redundancy, which make higher-order feature analysis possible. 
Based on the valid feature combination scores obtained from the above process, we select the top important ones and construct them as new regular features in the embedding layer to help the model better capture interaction patterns. We conducted extensive experiments on classical base models across three datasets to validate the effectiveness of our method.
The results clearly show that TayFCS improves prediction accuracy, while also demonstrating high efficiency in the analysis process and strong potential for real-world deployment, which may provide a practical and novel perspective for optimizing deep recommender systems.

\section*{Limitations}
Although TayFCS efficiently selects useful feature combinations and these combinations are model-agnostic—bringing improvements in prediction accuracy with acceptable inference latency—the additional combinations require extra embedding space, leading to increased memory cost. 
In fact, there is a lot of redundancy in the feature values of the constructed feature combinations, which is not addressed in TayFCS.
In future work, we aim to explore combinations at the feature value level to further enhance accuracy while minimizing memory overhead.

\bibliographystyle{ACM-Reference-Format}
\bibliography{sample-base}

\appendix
\section{The $T_3$ Term}\label{sec:t3}

The third-order term $T_3$ of the Taylor expansion is given by summing over all possible combinations of feature fields. The complete expression is:
\begin{equation}
\begin{aligned}
T_3 = \frac{1}{6} \bigg[ 
& \sum_{p} \frac{\partial^3 f(\boldsymbol{x}_0)}{\partial x_p^3} (\Delta x_p)^3 \\
& + 6 \sum_{\substack{i<j<k}} \frac{\partial^3 f(\boldsymbol{x}_0)}{\partial x_i \partial x_j \partial x_k} \Delta x_i \Delta x_j \Delta x_k \\
& + 3 \sum_{p} \sum_{q \neq p} \frac{\partial^3 f(\boldsymbol{x}_0)}{\partial x_p^2 \partial x_q} (\Delta x_p)^2 \Delta x_q 
\bigg],
\end{aligned}
\label{eq:t3_general}
\end{equation}

where each index $p, q, i, j, k$ refer to a feature field, and the second sum runs over all distinct ordered triples $i < j < k$.
Each term $\frac{\partial^{3} f(\boldsymbol{x}_0)}{\partial x_i \partial x_j \partial x_k} \Delta x_i \Delta x_j \Delta x_k$ corresponds to a third-order feature importance for a specific combination of three features. We use the proposed approximation method to efficiently calculate these derivatives for all possible feature combinations.

\section{Symbols \& Terminologies}
\label{sec:symbol}
\textbf{Field}
refers to the name of a data attribute, such as \textit{gender}. It represents a specific category or ``column" of the data.

\noindent\textbf{Feature}
refers to the value of a \textit{field}. For example, in the \textit{gender} field, features can be \textit{Male}, \textit{Female}, or other values.

\noindent\textbf{Feature Combination} is synthesized and crafted feature. For example, the combination of original features A, B, and C is feature ABC. After passing through the embedding layer, the encoding of ABC is on par with, and independent of, the encodings of A, B, and C.

\noindent\textbf{Order of Feature Combination} refers to the number of feature fields it involves. For example, feature combination ABC is a third-order feature. It can also be written as 3\textsuperscript{rd}-order or 3-order feature.

\noindent\textbf{Feature Interaction} refers to the effects and relationships between different features during the signal forward process in a model. For example, linear transformations of different features by each layer in a DNN are a common type of feature interaction.

\noindent\textbf{CVR} is the Conversion Rate. It measures the proportion of users who, after clicking on a recommended item (\textit{e.g.}, product or service), complete a specific goal (\textit{e.g.}, purchase, registration, download)~\cite{cvr}.

\end{document}